\documentclass[conference]{IEEEtran}
\usepackage{spconf,amsmath,graphicx}
\usepackage{amssymb}
\makeatletter

\usepackage{lineno}
\usepackage{epsfig}
\usepackage{cite}
\usepackage[utf8]{inputenc}
\usepackage{times}
\usepackage[figuresright]{rotating}
\usepackage{float}
\usepackage{amssymb}
\usepackage{algorithm}
\usepackage{algorithmic}
\renewcommand{\ALG@name}{Transceiver Design}
\usepackage{url}
\usepackage{hyperref}
\usepackage{flushend}

\title{High rate-reliability beamformer design for $2\times2$ MIMO-OFDM system under hostile jamming}
\name{Anu Jagannath$^{\dagger\ddagger}$, Jithin Jagannath$^\dagger$, Andrew Drozd$^\dagger$\thanks{ACKNOWLEDGMENT OF SUPPORT AND DISCLAIMER: (a) This material is based upon work supported by the US Army Contract No. W56HZV-19-C-0061. (b) Any opinions, findings and conclusions or recommendation expressed in this material are those of the author(s) and do not necessarily reflect the views of US Army.}}
\address{$^\dagger$ Marconi-Rosenblatt Innovation Laboratory, ANDRO Computational Solutions, LLC, Rome, NY\\
$^\ddagger$ Northeastern University, Boston MA, E-mail: {jagannath.a}@husky.neu.edu\\
E-mail: \{ajagannath, jjagannath, adrozd\}@androcs.com
}

\IEEEoverridecommandlockouts
\begin{document}
\newcommand{\squeezeupppp}{\vspace{-8 mm}}
\newcommand{\squeezeuppp}{\vspace{-6 mm}}
\newcommand{\squeezeupp}{\vspace{-5 mm}}
\newcommand{\squeezeup}{\vspace{-3 mm}}
\newcommand{\squeezeu}{\vspace{-2 mm}}
\newcommand{\squeeze}{\vspace{-1 mm}}
\newcommand{\squeez}{\vspace{-.5 mm}}
%
\maketitle
\begin{abstract}
Multiple-input multiple-output (MIMO) systems find immense potential and applicability in the long term evolution (LTE), 5G, Internet of Things (IoT), vehicular ad hoc networks (VANETs), and tactical communication systems. Jamming poses significant communication hindrance as well as security risks to the wireless communication systems. The achievable rate and reliability are the two most compromised aspects of a wireless link under such severe jamming. Owing to the high capacity and reliability of MIMO systems, they are increasingly used in tactical and critical applications. Therefore, it becomes essential to assess and enhance their sustenance under hostile jamming scenarios. To this end, we address the rate and reliability requirements of a MIMO OFDM system and propose a novel rate-reliability beamformer transceiver design for highly reliable and spectrally efficient operation under the most detrimental jamming attacks. We consider the disguised all band and multiband jamming where the jammer continuously attempts to mimic the legit transmissions. Additionally, we evaluate the rate and reliability performance under barrage jamming.

The significant contributions of the proposed rate-reliability beamformer scheme are: (i) achieves a minimum of 2 orders of magnitude better reliability in contrast to the state-of-the-art, (ii) outperforms the state-of-the-art scheme by $1.4\times$ with regards to achievable spectral efficiency, (iii) a very low complexity ($\mathcal{O}\left(\vert\mathcal{Q}\vert\right)$) decoder is presented, and (iv) first work to evaluate the performance of state-of-the-art transmit diversity scheme under hostile jamming attacks. 

\end{abstract}
\section{Introduction}
\label{sec:intro}

With the rise in the need for safety-critical vehicular and tactical communications, the need for jammer-aware or jammer resilient systems that can sustain communication links in the presence of jamming is increasing \cite{vanetattacks}. The spectrum efficiency and capacity of long term evolution (LTE) has led to its adoption for public safety, emergency-response, vehicular, and tactical communication systems \cite{tactLTE}. Tactical wireless networks are often subject to hostile jamming attacks. \emph{High reliability and higher spectral efficiency are the key requirements of such wireless links that are subject to jamming attacks.} These key requirements form the motivation and focus of this work. Jamming is a detrimental Denial-of-Service (DoS) attack in wireless communications that disrupts cellular networks \cite{lteattacks}. The effects of jamming on public safety communications \cite{IDAD}, tactical communications \cite{aj_tactical}, cellular systems \cite{aj_cellular}, etc. is a widely studied topic. 

In DoS attacks, the malicious attacker attempts to keep the communication channel busy to deny communication to legitimate users. With advancements in technology, the jamming threats are evolving to cognitive, disguised, and stealthy strategies. Such jamming attacks render traditional frequency hopping schemes, low probability of detection schemes to counteract reactive jamming, etc. less reliable.  The increased data rate, link range without additional bandwidth or transmit power requirements are leading to the widespread adoption of multiple antenna diversity systems \cite{anu_mimo,mimo_3,jagannath2020breaking}. Multiple-input multiple-output (MIMO) systems can be exploited to obtain diversity and spatial multiplexing gains leading to increased capacity. Antenna diversity provides the receiver with multiple copies of the transmitted symbols which enhances link reliability by combating fading and interference. Owing to the increased capacity of MIMO systems, several modern mobile wireless systems such as LTE, 4G, WiMAX, WiFi, HSPA+, and DVB-NGH have adopted multiple antenna diversity \cite{mimo_3}.  
Orthogonal Frequency Division Multiplexing (OFDM) is a popular scheme for broadband wireless communication (WLAN, cellular systems) that is robust against multipath fading. However, OFDM systems are severely affected by jamming. The effects of jamming attacks on OFDM systems was elaborately studied in \cite{ofdm_jam_1,ofdm_jam_2,ofdm_jam_3}. 

The increasingly severe hostile jamming attacks prompt the need to adopt jammer-resilient strategies. In this work, we address the rate and reliability aspects of a communication system to design a novel rate-reliability time-varying beamformer along with a very low complexity decoder for symbol detection at the receiver. The system under consideration here is a MIMO-OFDM communication system. 

This article is organized as; few related works are compared and contrasted in Section \ref{sec:rel}, Section \ref{sec:sys} elaborates the proposed rate-reliability beamformer, low complexity decoder, and the jamming attack model, Section \ref{sec:perf} describes the benchmark scheme and extensive performance evaluations, and finally the work is summarized in Section \ref{sec:con}.

\section{Related Works}
\label{sec:rel}

Jamming mitigation strategies for single and multiple antennas are a subject of active research. \cite{ofdm_prec,ofdm_multitone} proposed anti-jam transmit precoding strategies for single-antenna systems. The authors of \cite{siso_aj_fh} proposed a jammer-resistant shared key establishment using uncoordinated frequency hopping thereby breaking the circular dependency with preshared code. The shared key is then utilized to generate a frequency hopping pattern allowing the nodes to communicate in a coordinated frequency hopping manner in the presence of a jammer. Similarly, other jammer resistant uncoordinated frequency hopping schemes were analyzed in \cite{siso_ufh_2,siso_ufh_3,siso_ufh_4}. Han et.al \cite{OFDM_NBJam_Han} proposed a jammed pilot mitigation strategy for OFDM systems to detect and extract the pilot tones. The authors considered narrowband jamming where only the pilot tones are jammed. However, these schemes either consider the possibility of being able to switch frequencies in a given bandwidth or a narrowband jammer where only a portion of the spectrum is affected. Such schemes will fail for an all band disguised/barrage jamming where the entire signal spectrum or wider bandwidth (in case of barrage) is jammed.

Several works studied the benefit of combining MIMO with OFDM for wireless communications. The jamming resilience of the MIMO-OFDM system with interference cancellation and precoding was studied in \cite{mimo_ofdm_1}. 
A MIMO based blind jamming mitigation algorithm that leverages information from pilot bits was introduced in \cite{mimo_aj}. Their work is limited by the requirement of more antennas than the jammer at the receiver. Both these schemes focus on receive side mitigation strategy only. In contrast, our work proposes the rate improvement with transmit diversity scheme (essential to attain higher spectral efficiency) as well as a low complexity decoder to reliably detect the information symbols. A transmit-side beamforming for jammer mitigation was studied in \cite{mimo_aj_bf}. In this work, the authors considered a Gaussian distributed jamming signal which could be more representative of a barrage jammer. In our work, however, we consider disguised jamming under all and multi-band scenarios in addition to barrage where either all or a portion of the OFDM subcarriers are jammed. Furthermore, the disguised jamming represents the jamming attacks whereby the jammers mimic the legit user's transmissions making them an even harder detection problem at the receiver. Distributed MIMO decoding algorithms for decoding under jamming was proposed in \cite{mimo_aj_2}. A major limitation of this approach lies in the polynomial complexity of the decoder which grows as a \emph{polynomial} in the constellation size ($\vert\mathcal{Q}\vert$). Further, the authors do not reveal the spectral coverage of the jammer considered in their simulations. In contrast, our rate-reliability beamformer assures both higher rate and reliability under the most detrimental jamming attacks while achieving \emph{very low decoding complexity of $\mathcal{O}\left(\vert\mathcal{Q}\vert\right)$}. The low decoding complexity makes it suitable for practical implementation and commercial use.

Several works have studied the benefits of combining space time block coding (STBC) with beamforming \cite{STBC_BF, STBC_BF2, STBC_BF3_Pascual, STBC_BF4_Lin, STBC_BF5_Jongren, STBC_BF6_Cheng, STBC_BF7_Guan}. Beamforming can be adopted either in the transmit or receive side. The works in \cite{STBC_BF4_Lin, STBC_BF5_Jongren, STBC_BF7_Guan} studied transmit diversity scheme by combining the traditional Alamouti \cite{Alamouti} scheme with linear beamforming. However, a major limitation of this system in contrast to the proposed scheme is its symbol transmission rate and robustness to various jammer attacks. \textit{The performances of such transmit diversity schemes have not been demonstrated before under severe jamming.} The availability of portable jamming devices as well as the increased adoption of MIMO systems for tactical applications necessitates the need to study the effects of jamming on these conventional schemes. To this end, we will conduct extensive numerical simulations in Section \ref{sec:perf} to compare and contrast the effects of jamming on a state-of-the-art scheme and the proposed rate-reliability beamformer. This work aims to improve the resiliency of MIMO systems against jamming attacks with the following contributions:\begin{itemize}
\item In order to combat jamming, we propose a novel high rate-reliability time-varying beamformer design. \emph{This is the first work that unifies achievable rate as well as reliability in the beamformer design for operation under disguised hostile jamming.}
\item In order to achieve a higher rate and reliability without increasing the complexity, we present the conditional ML decoding of complexity $\mathcal{O}\left(\vert \mathcal{Q}\vert\right)$ to decode the information symbols from the beamformed transmissions.
\item A highly reliable, high rate MIMO design that is robust to different types of jamming attacks are presented and demonstrated with extensive numerical simulations. 
\item First work to study the effects of various hostile jamming attacks on widely adopted transmit diversity scheme.
\end{itemize}
\section{Problem Formulation}
\label{sec:sys}

In this section, we present a detailed account of the proposed rate-reliability beamforming strategy, a low complexity decoder, and the various jamming attacks considered in this work.
\subsection{Rate-Reliability Beamformer Design}
\label{sec:design}

The key requirements in our design are, 

\textbf{(i) High Rate}: The high rate will be achieved by exploiting the space-time diversity of STBC. Here, we consider a $2\times2$ MIMO system, the conventional Alamouti scheme offered a symbol rate of only $1$ symbols/s whereas the STBC proposed in \cite{rate2} offered a symbol transmission rate of $2$ symbols/s. In this design, therefore we will be considering this rate-2 STBC.

\textbf{(ii) High Reliability}: Conventional beamforming \cite{bf_3} improves the symbol reliability by transmitting all available power along the channel's strongest direction. Beamformer can be viewed as the precoding to the symbols from multiple transmit antennas along the intended channel direction. To further improve the resiliency, we will incorporate the minimum bit error rate (BER) precoding \cite{ofdm_prec} to the design.

We will incorporate the aforementioned ingredients to result in a rate-reliability beamformer design. Eigen-Beamforming is a MIMO technique whereby the system capacity is enhanced by transmitting multiple beams pointing to orthogonal directions along the eigenvectors of the channel’s correlation matrix. The proposed jammer resilient rate-reliability beamformer design can be expressed as 
\begin{equation}
    \mathbf{B}_{RR}=\mathbf{C}_{p}\mathbf{D}_{H}^{1/2} \mathbf{U}_{H}^H \label{eq:B2d}
\end{equation}
where $\mathbf{C}_p$ is the precoded $2\times2$ rate-2 STBC \cite{rate2} (with rotation angle $\phi$) for Quadrature Amplitude Modulation (QAM) modulations, $\mathbf{D}_{H}$ is the diagonal power loading matrix, and $\mathbf{U}_{H}= \left[\begin{matrix}u_{11} & u_{12}\\u_{21} &u_{22} \end{matrix}\right]$ is the eigenvector matrix whose columns $(\mathbf{u}_a,\mathbf{u}_b)$ are eigenvectors of channel correlation matrix $\mathbf{R}_{H}$. The spectral decomposition of $\mathbf{R}_{H}$ is $\mathbf{R}_{H}=\mathbf{U}_{H} \mathbf{D}_{H}^{1/2} \mathbf{U}_{H}^H$ where $\mathbf{D}_{H}=diag(\delta_1,\delta_2 )$ is the diagonal matrix with diagonal elements $\delta_1,\delta_2$. The MIMO transceiver adapts the power loading ($\mathbf{D}_{H}$) on the eigen-beams and the eigenvectors to direct the beams based on the estimated $2\times 2$ channel matrix $\mathbf{H}$. 

Precoding of the STBC can be performed after STBC encoding on the two output symbol streams from the STBC encoder. 
The STBC encodes four information symbols $\{x_1, x_2, x_3, x_4\}\in \mathcal{Q}$ over two transmit antennas and over two epochs yielding a rate-2 transmission ($\mathcal{R}=2$symbols/s/Hz). The STBC ($\mathbf{C}$) takes the form,
\begin{equation}
    \mathbf{C} = \begin{bmatrix}x_1\sin{\phi_1}-x_2^*\cos{\phi_1} & x_3\sin{\phi_2}-x_4^*\cos{\phi_2}\\-x_3^*\sin{\phi_2}+x_4\cos{\phi_2} &x_1^*\sin{\phi_1}-x_2\cos{\phi_1} \end{bmatrix}\label{eq:stbcc}
\end{equation}
 For notational simplicity, let us represent the $2\times2$ STBC in equation (\ref{eq:stbcc}) as, 
\begin{equation}
    \mathbf{C} = \begin{bmatrix}C_1 & C_2 \\ -C_2^* & C_1^* \end{bmatrix}\label{eq:stbc}
\end{equation}
where $C_1 = x_1\sin{\phi_1}-x_2^*\cos{\phi_1}$ and $C_2 = x_3\sin{\phi_2}-x_4^*\cos{\phi_2}$. The optimal angles $\phi_1$ and $\phi_2 = \frac{\pi}{2}-\phi_1$ are chosen to maximize the diversity and coding gain as in \cite{rate2}. Here, the rows and columns of $\mathbf{C}$ represent the channel use (epochs/time slots) and transmit antennas respectively. Hence, $4N$ QAM information symbols will yield two symbol vectors $\mathbf{s}_i\in \mathbb{C}^{2N}$ at the output of STBC encoder. The precoded STBC can be given as
\begin{equation}
    \mathbf{C}_p =  \mathbf{P}\begin{bmatrix}\mathbf{s}_1 & \mathbf{s}_2\end{bmatrix}
\end{equation}
where $\mathbf{P}$ is the precoder matrix and $N_c$ is the number of OFDM subcarriers. Additionally, $\sum_k \rho_k^2 = \mathcal{P}$ imposes a total power constraint to the precoding strategy. Two types of precoding are considered; full and multi-band precoding. As the names suggest, by definition a full precoder protects the entire OFDM subcarriers whereas the multi-band precoder protects an arbitrary set of subcarriers. Accordingly, the precoder matrices for the full and multi-band precoders are defined as
\begin{align}
    \mathbf{P}_{F} =& \begin{bmatrix} diag\left(\rho_1,\rho_2,\cdots,\rho_{2N} \right)\\ \mathbf{0}
\end{bmatrix}_{N_c\times 2N},\\  \mathbf{P}_{M} =& \begin{bmatrix} diag\left(\boldsymbol{\rho} \right)\\ \mathbf{0}
\end{bmatrix}_{N_c\times 2N}
\end{align}
where is $\boldsymbol{\rho}\in \mathbb{R}^{2N}$ is a vector with non-zero values at jammed subcarrier locations. The rate-reliability beamformer of equation (\ref{eq:B2d}) can be expanded as in equation (\ref{eq:BRR}).
\begin{figure*}[!t]
\begin{equation}
   \mathbf{B}_{RR} = \begin{bmatrix}
    \sqrt{\delta_1}C_{1_p} u_{11}^* + \sqrt{\delta_2}C_{2_p} u_{12}^* &\sqrt{\delta_1}C_{1_p} u_{21}^* + \sqrt{\delta_2}C_{2_p} u_{22}^*\\
-\sqrt{\delta_1}C_{2_p} u_{11}^* + \sqrt{\delta_2}C_{1_p} u_{12}^* &-\sqrt{\delta_1}C_{2_p} u_{21}^* + \sqrt{\delta_2}C_{1_p} u_{22}^*
    \end{bmatrix}\label{eq:BRR}
\end{equation}
\end{figure*}
Here, $C_{1_p},C_{2_p}$ simply denote the precoded STBC symbol encodings as introduced in the previous section. Hence, our beamformer $\mathbf{B}_{RR}$ can be viewed as a time-varying beamformer with the rows acting as beam-steering vectors at each epoch. The rows of the beamformer are projected on $2$ orthogonal eigenvectors that are invariant with respect to the $2$ time slots. Similarly, the power loading factors dictated by the constants $\delta_1$ and $\delta_2$ are time-invariant across the 2 epochs. Hence, the beamformer $\mathbf{B}_{RR}$ allocates invariant power along invariant eigen directions that are multiplexed with different weights (time and transmit antenna variant) as dictated by the elements of the precoded STBC $\mathbf{C}_p$. The optimal eigen-beams are power-loaded according to a spatial water-filling principle. In this way, we have designed a rate-reliability beamformer design that enjoys the benefits of high spectral efficiency and reliability.
\subsection{Decoder Design}
\label{sec:decoderdesign}
Assuring higher rate and reliability under hostile jamming attacks while maintaining a low decoder complexity is essential to promote practical use. To this end, we design a decoder with very low decoding complexity. Consider a coherent MIMO system with the received signal model expressed as,
\begin{equation}
    \mathbf{Y} = \sqrt{\mathcal{P}}\mathbf{B}_{RR}\mathbf{H} + \mathbf{J} + \mathbf{N} \label{eq:Y}
\end{equation}
where $\mathcal{P}$ is the transmit power per antenna,$\mathbf{Y}=\begin{bmatrix}\mathbf{y}_1 & \mathbf{y}_2\end{bmatrix}$ is the received signal matrix, $\mathbf{y}_i = \begin{bmatrix} y_i^1 & y_i^2\end{bmatrix}^T$ is the received signal vector at the $i$-th receiver antenna, $\mathbf{B}_{RR}$ is the rate-reliability beamformer  matrix, $\mathbf{H}$ is the Rayleigh flat fading channel matrix, $\mathbf{J}$ is the hostile disguised jamming signal, and $\mathbf{N}$ is the additive white Gaussian noise matrix. The superscript $1$ and $2$ indicates the epochs. The first step at the receiver, post filtering,synchronization, and OFDM demodulation is the full/multi-band decoding as given by
\begin{equation}
    \hat{\mathbf{C}} = \mathbf{D}\mathbf{Y} \label{eq:fmdec}
\end{equation}
where $\mathbf{D}$ is the full/multi-band decoder expressed as 
\begin{align}
    \mathbf{D}_F =&  \begin{bmatrix}
diag\left(1/\rho_1,1/\rho_2,\cdots,1/\rho_K \right) & \mathbf{0}
\end{bmatrix}_{2N\times N_c} \\
\mathbf{D}_M =& \begin{bmatrix}
diag\left(\boldsymbol{1/\rho}\right) & \mathbf{0}
\end{bmatrix}_{2N\times N_c}
\end{align}
Here, $\boldsymbol{1/\rho}$ is the vector whose elements are $1/\rho_i$ at the jammed subcarrier locations.
 Considering the channel matrix $\mathbf{H} = \left[\begin{matrix} \mathbf{h}_1 & \mathbf{h}_2\end{matrix}\right]= \left[\begin{matrix}h_{11} &h_{12}\\h_{21} &h_{22} \end{matrix}\right]$, the received symbols at $i$-th receiver antenna during time slot-1 can be expressed as in equation (\ref{eq:yi0}).
 \begin{figure*}[!t]
\begin{align}
    y_i^1 =& 
    \left(\sqrt{\delta_1}u_{11}^*h_{1i} + \sqrt{\delta_1}u_{21}^*h_{2i}\right)\hat{C}_1 + 
    \left(\sqrt{\delta_2}u_{12}^*h_{1i} + \sqrt{\delta_2}u_{22}^*h_{2i} \right)\hat{C}_2 + j_i^1 + n_i^1 \label{eq:yi0}
\end{align}
\end{figure*}
Similarly, the conjugate of the expression corresponding to time slot-2 can be represented as in equation (\ref{eq:yi1}).
\begin{figure*}[!t]
\begin{align}
    y_i^{2^*} =& \left(\sqrt{\delta_2}u_{12}^*h_{1i} + \sqrt{\delta_2}u_{22}^*h_{2i}\right)\hat{C}_1 -
    \left(\sqrt{\delta_1}u_{11}^*h_{1i} + \sqrt{\delta_1}u_{21}^*h_{2i} \right)\hat{C}_2 + j_i^{2^*} +n_i^{2^*} \label{eq:yi1}
\end{align}
\end{figure*}
In equations (\ref{eq:yi0}) and (\ref{eq:yi1}), $j_i^1$, $j_i^2$, $n_i^1$ and $n_i^2$ represent the jammer and noise terms respectively. The notations $\hat{C}_1, \hat{C}_2$ imply the detected equivalents of the STBC elements. Expressing $\mathbf{y}_i$ in the equivalent virtual channel fading matrix (EVCM) manner we get equation (\ref{eq:evcm}),
\begin{figure*}[!t]
\begin{align}
\label{eq:evcm}
    &\begin{bmatrix}y_i^1\\y_i^{2^*} \end{bmatrix} = \begin{bmatrix}j_i^1\\j_i^{2^*}\end{bmatrix}+\begin{bmatrix}n_i^1\\n_i^{2^*}\end{bmatrix}+ \begin{bmatrix}\hat{C}_1\\\hat{C}_2 \end{bmatrix}
    \underbrace{\begin{bmatrix}\left(\sqrt{\delta_1}u_{11}^*h_{1i} + \sqrt{\delta_1}u_{21}^*h_{2i}\right) &\left(\sqrt{\delta_2}u_{12}^*h_{1i} + \sqrt{\delta_2}u_{22}^*h_{2i} \right)\\\left(\sqrt{\delta_2}u_{12}^*h_{1i} + \sqrt{\delta_2}u_{22}^*h_{2i}\right) &-\left(\sqrt{\delta_1}u_{11}^*h_{1i} + \sqrt{\delta_1}u_{21}^*h_{2i} \right)  \end{bmatrix}}_{\mathbf{G}_i}
\end{align}
\end{figure*}
where $\mathbf{G}_i$ is the EVCM with respect to the $i$-th receiver antenna. The equivalent form after channel equalization is expressed in equation (\ref{eq:Brreq}).
\begin{figure*}[!t]
\begin{align}
\label{eq:Brreq}
\begin{bmatrix}a_{i}^{1}\\a_{i}^{2} \end{bmatrix} = \left( \delta_1\|\mathbf{h}_i\mathbf{u}_{a}\|^2 + \delta_2\|\mathbf{h}_i\mathbf{u}_b\|^2\right)\mathbf{I}_2\begin{bmatrix}\hat{C}_1\\\hat{C}_2 \end{bmatrix} + \mathbf{G}_i^H\begin{bmatrix}j_i^1\\j_i^{2^*} \end{bmatrix} +\mathbf{G}_i^H\begin{bmatrix}n_i^1\\n_i^{2^*} \end{bmatrix}
\end{align}
\end{figure*}
The conditional maximum likelihood (ML) decoding can be now performed on the received symbols from the sufficient statistic and intermediate symbols. The sufficient statistics are obtained from the below equation
\begin{align}
    r^j = \frac{1}{2}\sum_{i=0}^{1}a_i^j 
\end{align} where $j = 1,2$ implies the time slot notation, $j=1$ will give the symbols $x_1,x_2$ and $j=2$ will yield $x_3,x_4$. The intermediate signals can be represented as
\begin{equation}
    \Tilde{r}^j = r^j - \sqrt{\frac{\mathcal{P}}{4}}\Psi\left[ -x_{2i}^*\cos{\phi_i} \right] \label{eq:inter}
\end{equation}
where $\Psi$ is the diagonal element of $\mathbf{G}_i\mathbf{G}_i^H$. For each of the $Q$ constellation points, the conditional ML estimate $x_{2i-1 | 2i}$ is obtained by minimizing the cost function
\begin{equation}
    \Lambda\left( x_{2i-1 | 2i}\right) = \left|r^j - \sqrt{\frac{\mathcal{P}}{4}}\Psi\left[x_{2i-1|2i}\sin{\phi}-x_{2i}^*\cos{\phi}\right]\right|^2.\label{eq:cost}
\end{equation}
The symbol pair $x_{2i-1 | 2i}, x_{2i}$ that minimizes the above cost function yields the correctly decoded pair. The complexity of the presented conditional ML decoding is $\mathcal{O}\left(\vert\mathcal{Q}\vert\right)$. The summary of encoder and decoder steps to achieve the rate-reliability beamforming is shown in Transceiver Design \ref{alg:algo}. 
\begin{algorithm}
\caption{Rate-Reliability Beamformer}
 \begin{algorithmic}[1]
 \renewcommand{\algorithmicrequire}{\textbf{Input:}}
 \renewcommand{\algorithmicensure}{\textbf{Output:}}
 \REQUIRE QAM Information symbols $\{x_1, x_2, \cdots \} \in \mathcal{Q}$
 \ENSURE  Decoded symbols $\{\hat{x}_1, \hat{x}_2, \cdots \}$\\ \textbf{Transmitter}:\\
\STATE \text{Rate-2 STBC Encoder}:\\ 
 $\mathbf{C} = \begin{bmatrix}x_1\sin{\phi_1}-x_2^*\cos{\phi_1} & x_3\sin{\phi_2}-x_4^*\cos{\phi_2}\\-x_3^*\sin{\phi_2}+x_4\cos{\phi_2} & x_1^*\sin{\phi_1}-x_2\cos{\phi_1} \end{bmatrix}$\\
 $4N$ information symbols yield two symbol vectors $\mathbf{s}_i\in \mathbb{C}^{2N}$\\
 \STATE \text{Full or Multi-band precoding}:
 \\$\mathbf{C}_p =  \mathbf{P}\begin{bmatrix}\mathbf{s}_1 & \mathbf{s}_2\end{bmatrix}$\\
 \STATE \text{OFDM Modulation}
\STATE  \text{Rate-Reliability beamformer}
\\$\mathbf{B}_{RR} =\mathbf{C}_{p}\mathbf{D}_{H}^{1/2} \mathbf{U}_{H}^H$
\\ \textbf{Receiver}:\\
\STATE \text{OFDM Demodulation}
\STATE \text{Full or Multi-band decoder Equation (\ref{eq:fmdec})}
\STATE \text{Channel Equalization} Equation (\ref{eq:Brreq})
\STATE \text{Obtain Intermediate Signals} Equation (\ref{eq:inter})
\STATE \text{Symbol pair detection by minimizing cost}\\ \text{function in Equation (\ref{eq:cost})}
 \RETURN $\{\hat{x}_1, \hat{x}_2, \cdots \}$
 \end{algorithmic}\label{alg:algo}
\end{algorithm}

\subsection{Jamming Attack Model}
Jamming is a major DoS attack by which the adversaries can disrupt the communication resulting in throughput degradation and network holes. In this work, we will consider two types of jamming: (i) Disguised and (ii) Barrage.
\begin{figure}[h!]
\centering
\epsfig{file=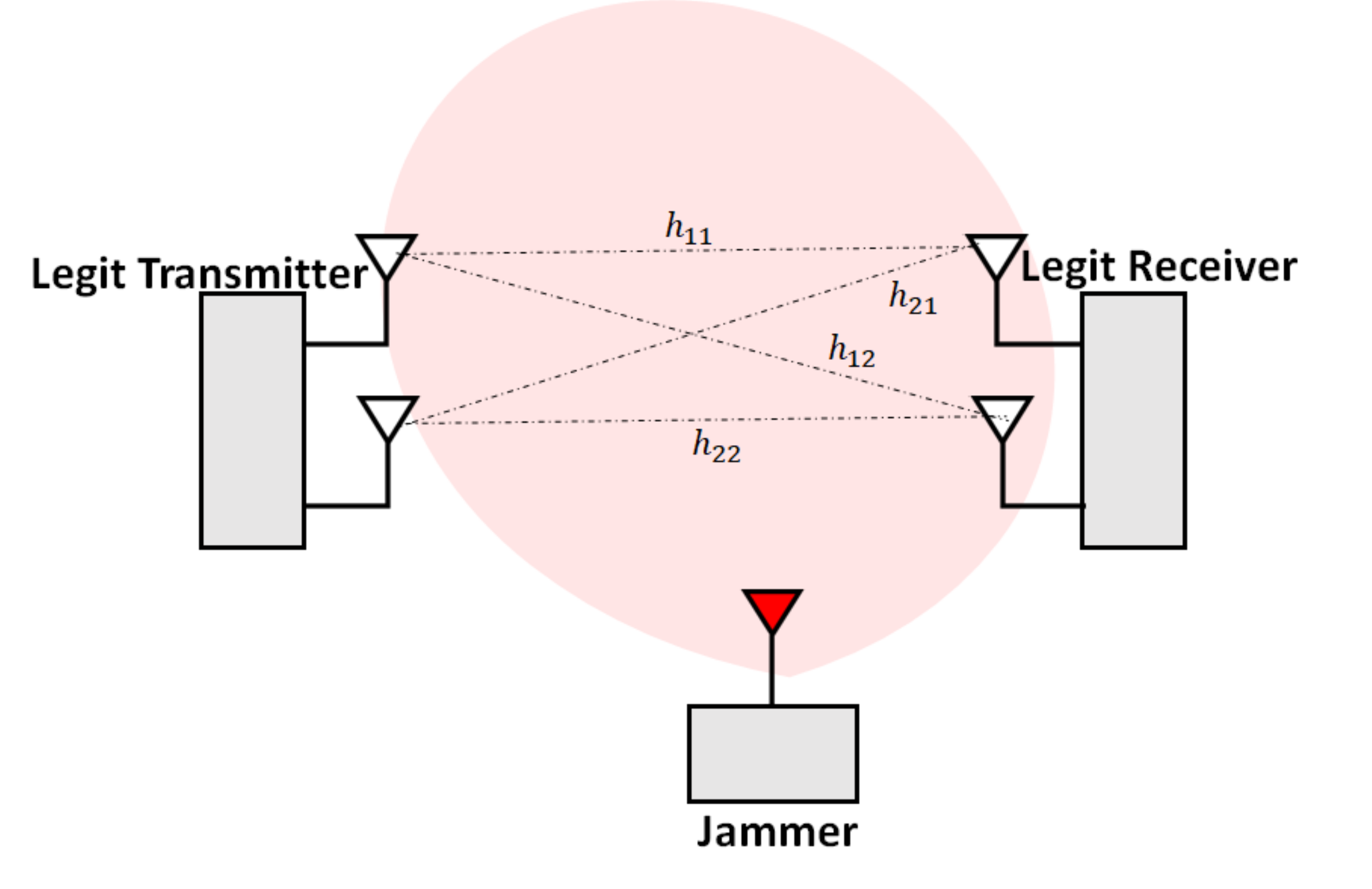, width=3 in,}
\caption{Legit Transmitter-Receiver $2\times2$ MIMO link under active jamming}\label{fig:setup}

\end{figure}

We define disguised jamming as the scenario where the jammer attempts to mimic the legit users by choosing the symbols from the same symbol constellation. In the OFDM system considered in this work, we define the disguised jammer to be following the same symbol modulation and OFDM structure as the legit user. This is an increasingly severe hostile environment where the similarity between the jammer and legit user makes it hard to distinguish at the receiver. We consider two types of disguised jammer: (i) All band and (ii) Multi-band. All band jammer covers the entire OFDM subcarriers as the legit user whereas the multi-band will only affect an arbitrarily chosen portion of the spectrum.

Barrage jamming refers to the high power white Gaussian noise transmission from the hostile transmitter. Such jamming attacks is distributed over a wide spectrum band. In all cases, \emph{the jammer is assumed to continuously transmit and therefore the legit user is always operating in the presence of active jamming.} The envisioned jamming scenario in which the legit user's communication link is under continuous jamming is depicted in Fig.\ref{fig:setup}.

\section{Performance Analysis}
\label{sec:perf}
\subsection{Conventional Alamouti-Beamformer}
\label{sec:BFalam}
State-of-the-art wireless communication standards such as IEEE802.11n, IEEE802.16e, LTE, and LTE-Advanced employ the Alamouti scheme to achieve higher spectral efficiency by exploiting transmit diversity. These communication standards are widely used in tactical, vehicular, industrial, and cellular applications. Hence, we will use the Alamouti beamformed systems \cite{STBC_BF4_Lin, STBC_BF5_Jongren, STBC_BF7_Guan} as the performance benchmark. In this section, we will discuss the conventional Alamouti beamforming technique as the preface to our performance evaluations. The conventional Alamouti based linear beamformer can be expressed by the 2D matrix, 
\begin{equation}
    \mathbf{B}_c = \mathbf{X}\mathbf{D}_{H}^{1/2} \mathbf{U}_{H}^H \label{eq:BF_a}
\end{equation}
where $\mathbf{X}$ is the $2\times2$ Alamouti STBC formed of information symbols $\{x_1, x_2\}$ given by $\mathbf{X} = \begin{bmatrix}x_1 & x_2 \\ -x_2^* & x_1^* \end{bmatrix}$.
Following the receiver model in equation (\ref{eq:Y}), the EVCM form can be derived as
\begin{equation}
    \begin{bmatrix}
    y_i^1 \\ y_i^{2*}
    \end{bmatrix} = \underbrace{\begin{bmatrix}
\sqrt{\delta_1}\mathbf{u}_a^H\mathbf{h}_i  &\sqrt{\delta_2}\mathbf{u}_b^H\mathbf{h}_i\\ \sqrt{\delta_2}\mathbf{u}_b\mathbf{h}_i^H & -\sqrt{\delta_1}\mathbf{u}_a\mathbf{h}_i^H    
    \end{bmatrix}}_{\mathbf{G}_i} \begin{bmatrix}
    x_1 \\ x_2
    \end{bmatrix} +\begin{bmatrix}
    j_i^1 \\ j_i^{2*}
    \end{bmatrix} + \begin{bmatrix}
    n_i^1 \\ n_i^{2*}
    \end{bmatrix}
\end{equation}
The symbol detection can be performed by the following linear operation at the receiver
\begin{equation}
     \begin{bmatrix}
    \hat{x}_1 \\ \hat{x}_2
    \end{bmatrix} = \mathbf{G}_i^H\begin{bmatrix}
    y_i^1 \\ y_i^{2*}
    \end{bmatrix}
\end{equation}
The single symbol decodability of the Alamouti scheme allows for symbol detection by linear processing at the receiver. The higher rate STBC inclusion in our rate-reliability beamformer design requires specific decoding at the receiver as detailed by the conditional ML decoding shown in section \ref{sec:decoderdesign}. However, the ability to incorporate the higher symbol transmission rate to the rate-reliability beamformer while maintaining a low decoding complexity of $\mathcal{O}\left(\vert\mathcal{Q}\vert\right)$ is a significant contribution of this work.

\begin{figure}[htb]
\centering
\epsfig{file=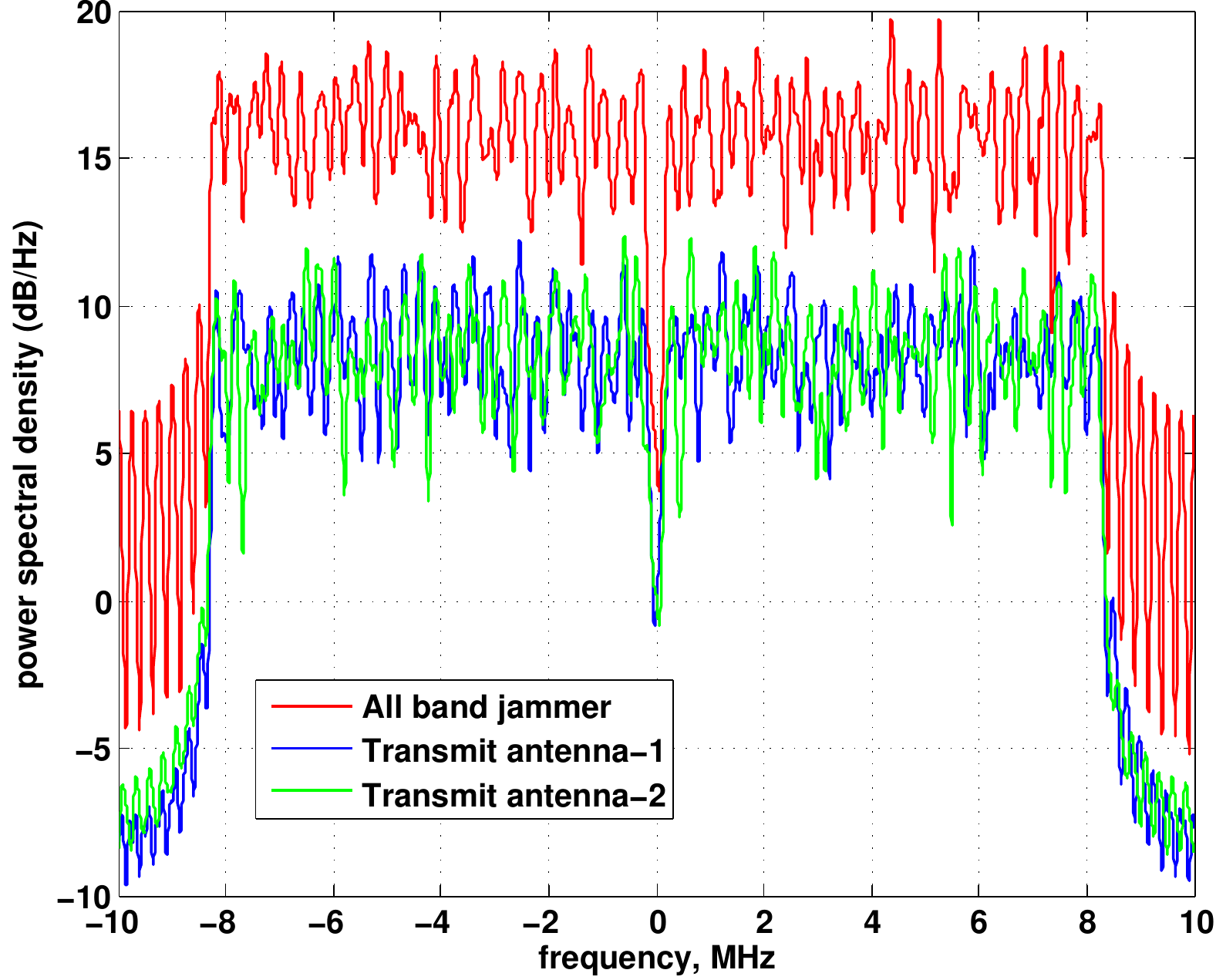, width=3 in,}
\caption{Power Spectrum Density under All band Jamming}\label{fig:allspec}
\end{figure}
\begin{figure}[htb]
\centering
\epsfig{file=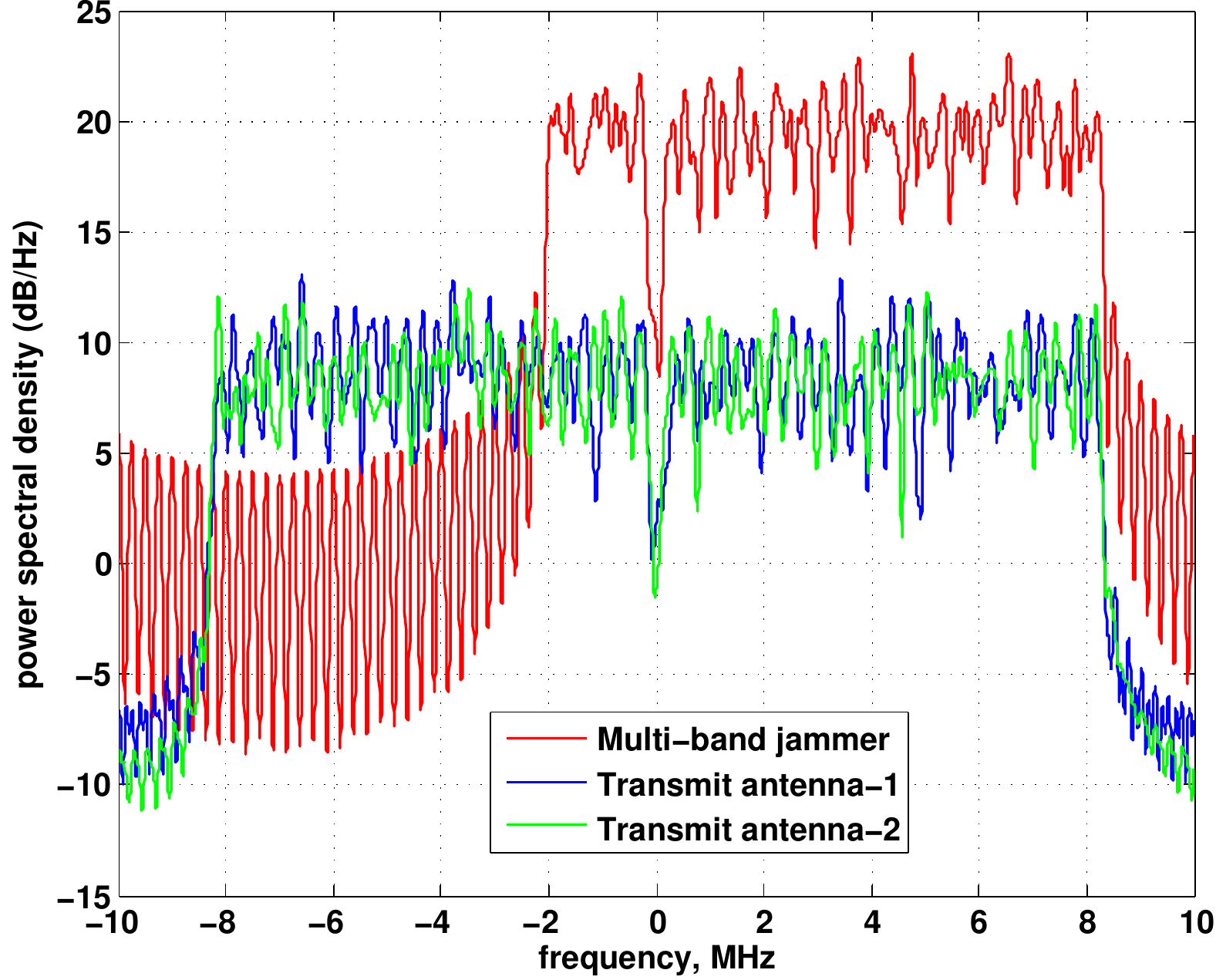, width=3 in,}
\caption{Power Spectrum Density under Multi-band Jamming}\label{fig:multispec}
\end{figure}
\begin{figure}[htb]
\centering
\epsfig{file=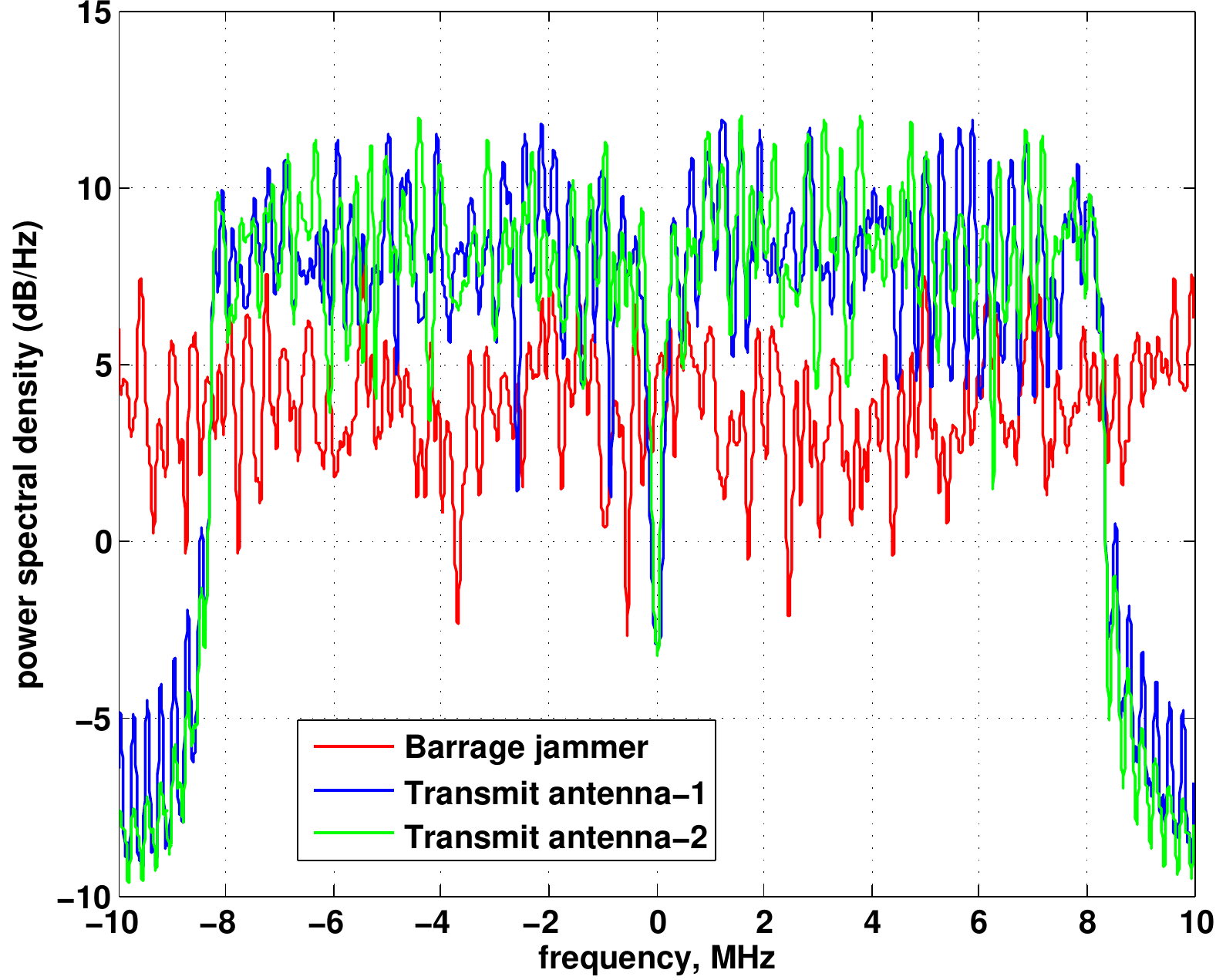, width=3 in,}
\caption{Power Spectrum Density under Barrage Jamming}\label{fig:barrspec}
\end{figure}

\subsection{Performance Evaluations}
We validate the performance of the proposed system design under diverse jammer threats in MATLAB simulations. The OFDM scheme is similar to IEEE802.11a which has 64 subcarriers of which 52 are data subcarriers. The simulations further consider flat-fading Rayleigh channel in AWGN and $4$-QAM modulation to achieve a maximum spectral efficiency of $4$ b/s/Hz whereas, with lower symbol transmission rate STBC such as Alamouti, the QAM constellation must be 16 to achieve the same spectral efficiency. The power spectral densities of the all band, multi-band, and barrage jammers are shown in Figures \ref{fig:allspec}, \ref{fig:multispec}, \ref{fig:barrspec}. Figure \ref{fig:allspec} shows the disguised all band jammer under very high jamming power which completely buries the legit user's signal. Figure \ref{fig:multispec} shows the disguised multi-band scenario where an arbitrary set of subcarriers are jammed and therefore a portion of the legit user's signal is buried in the jamming signal. Finally, the barrage jamming which sends Gaussian noise over a wide spectrum (wider than that of the legit user's signal) is depicted in Fig.\ref{fig:barrspec}. 

\begin{figure}[t!]
\centering
\epsfig{file=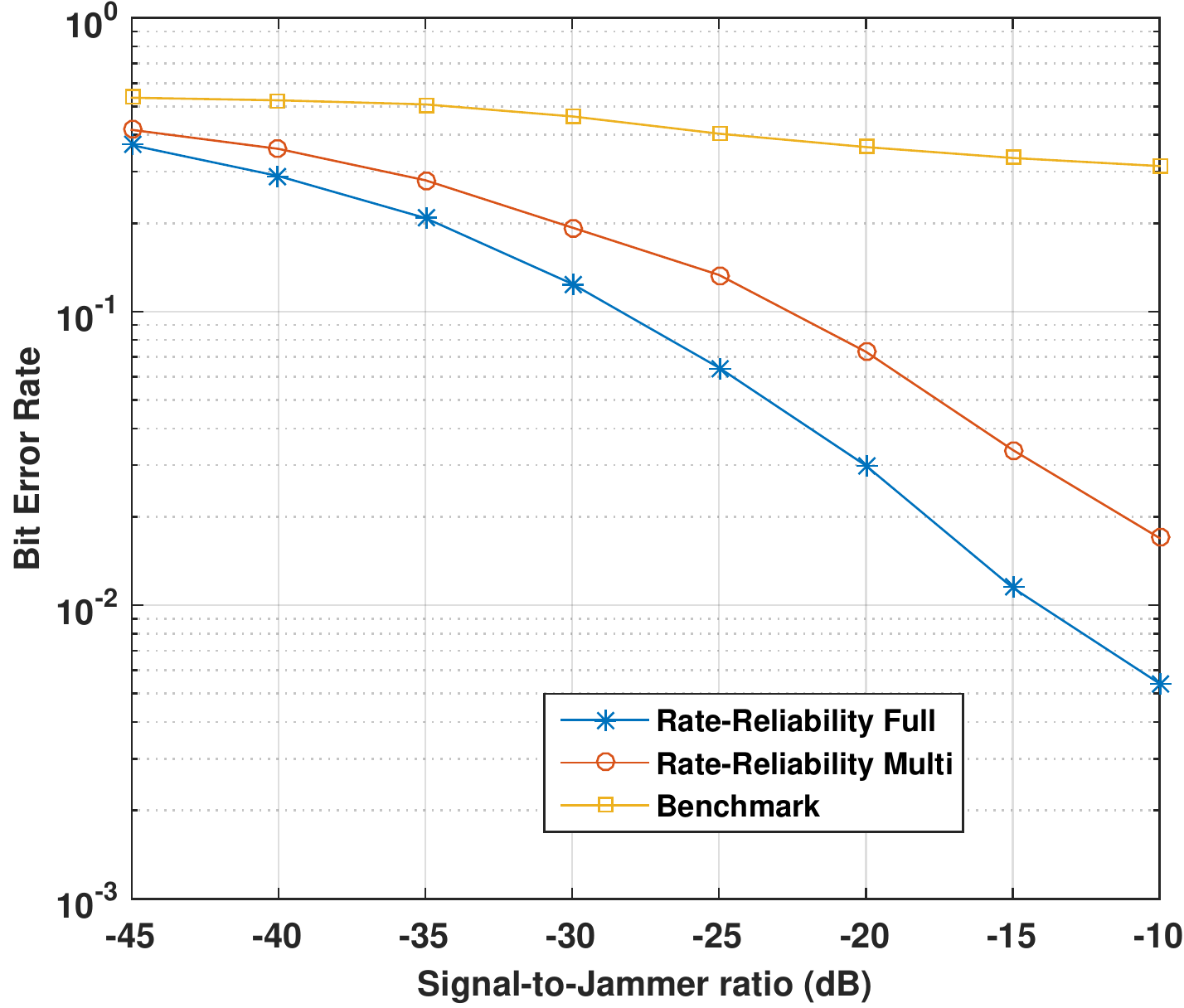, width=3 in,}
\caption{Bit error performance under all band jamming}\label{fig:fp_all}

\end{figure}

\begin{figure}[t!]
\centering
\epsfig{file=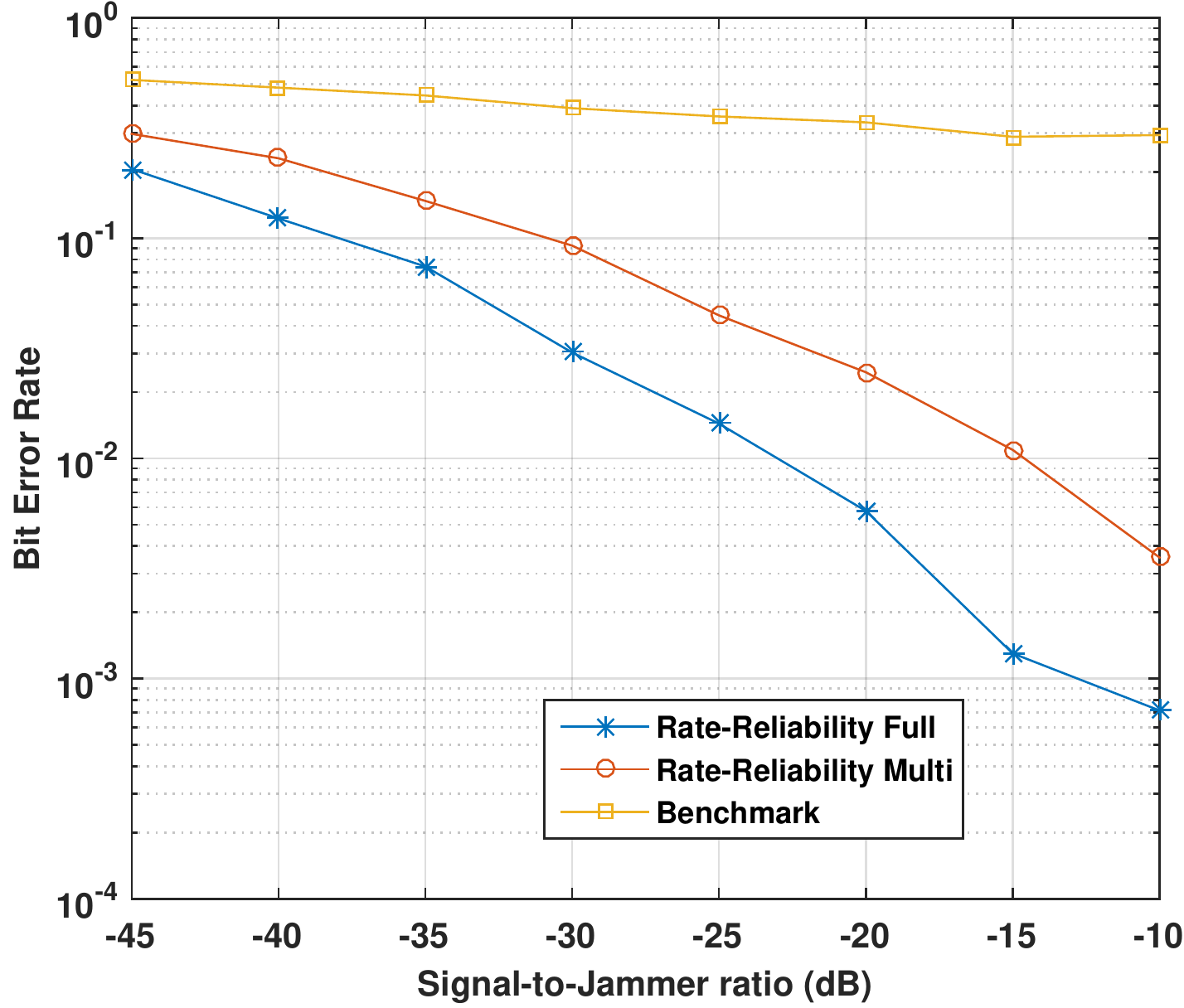, width=3 in,}
\caption{Bit error performance under barrage jamming}\label{fig:fp_barr}

\end{figure}

\begin{figure}[t!]
\centering
\epsfig{file=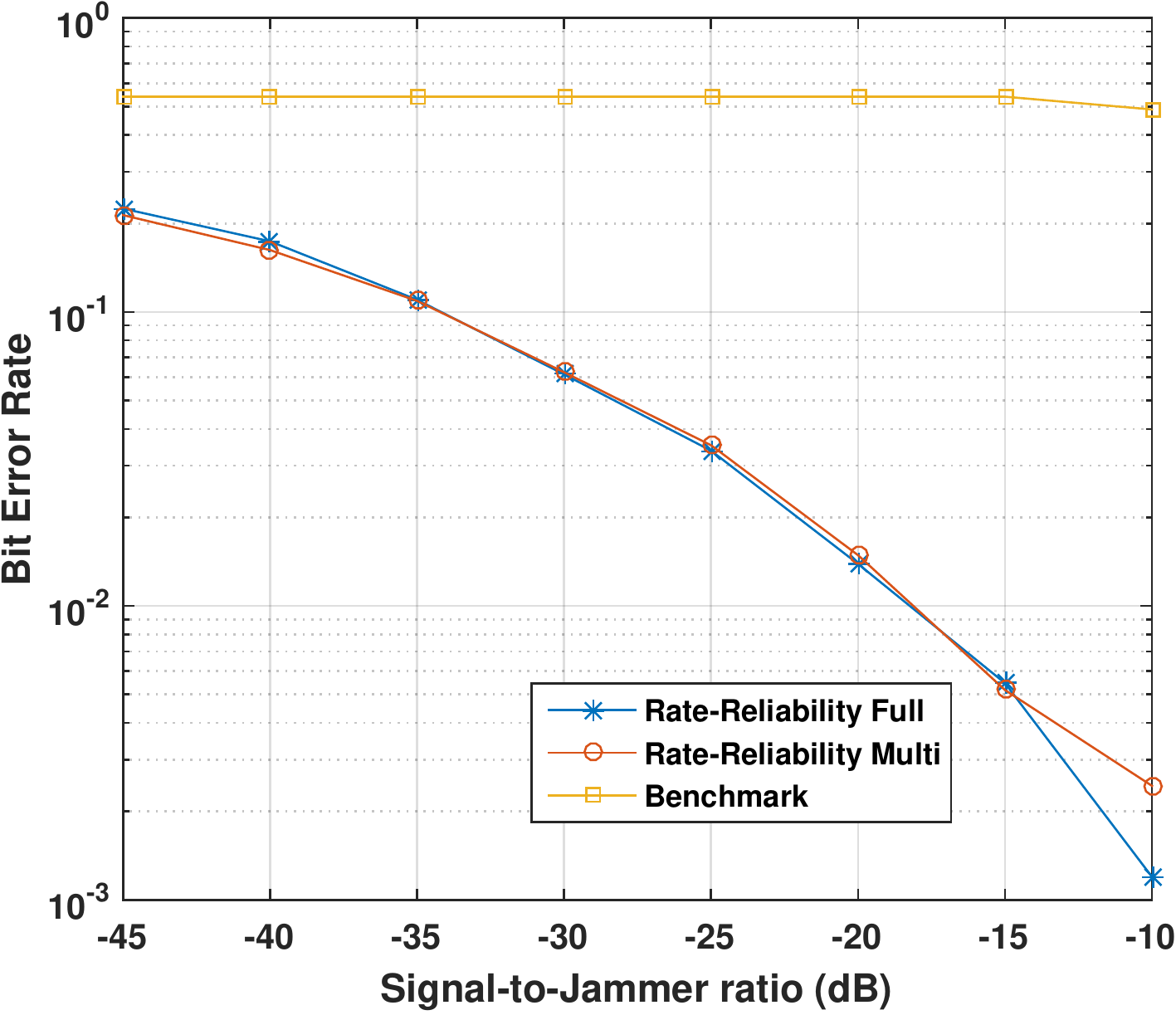, width=3 in,}
\caption{Bit error performance under multi-band jamming}\label{fig:fp_mb}

\end{figure}

\begin{figure}[t!]
\centering
\epsfig{file=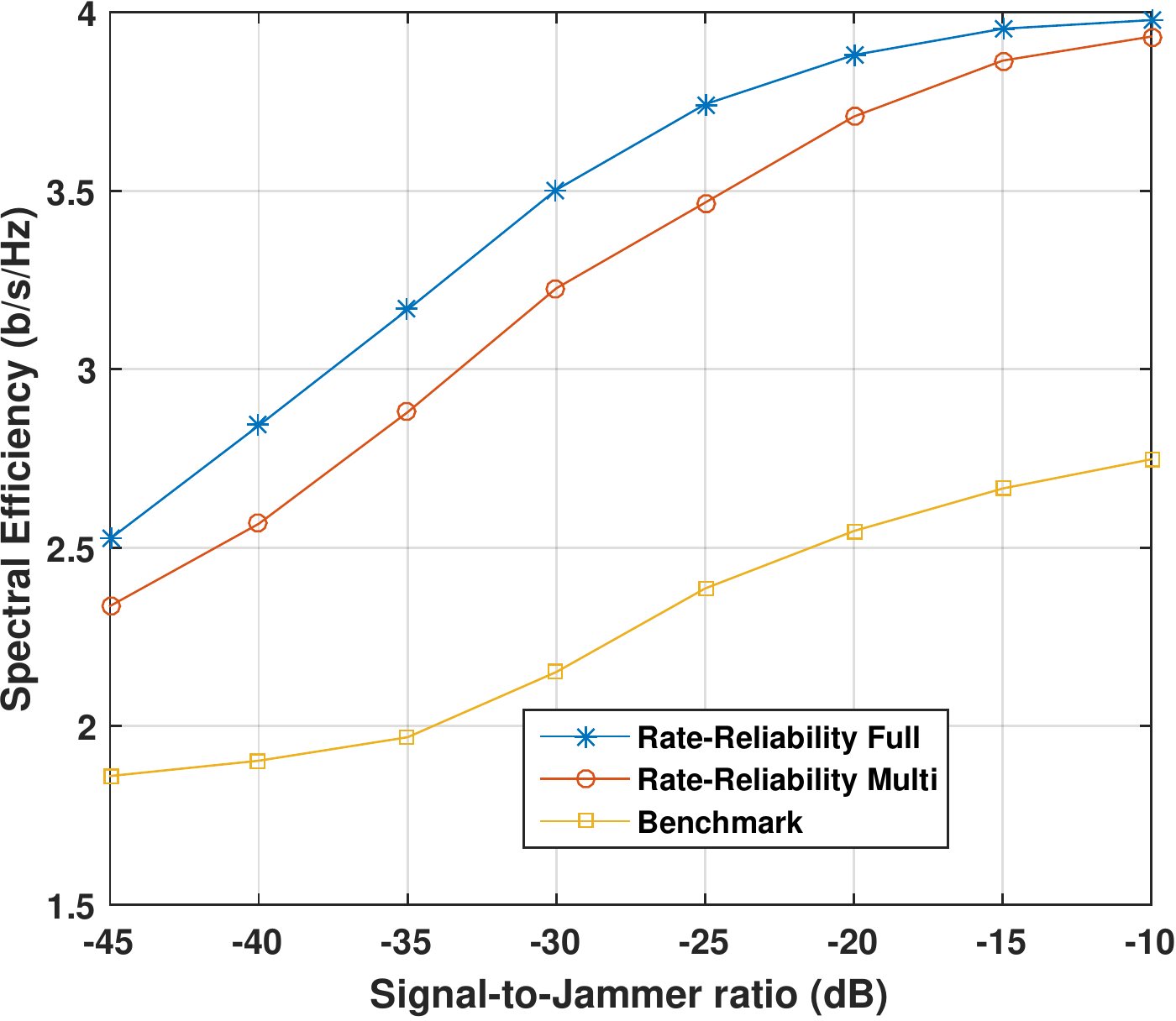, width=2.8 in,}
\caption{Spectral efficiency under all band jamming}\label{fig:mp_mb}

\end{figure}

In our simulations, we will analyze and contrast the performances of the proposed rate-reliability beamformer with full and multiprecoding as well as the conventional Alamouti beamformer detailed in section \ref{sec:BFalam}. The legends in the upcoming performance curves in Figures \ref{fig:fp_all}, \ref{fig:fp_barr}, \ref{fig:fp_mb}, and \ref{fig:mp_mb} correspond to the rate-reliability beamformer with full, multiprecoding, and the benchmark scheme (discussed in section \ref{sec:BFalam}) respectively. 

Figure \ref{fig:fp_all} denotes the disguised all band jamming scenario where the jamming power is increased from very high to low as denoted by the signal to jammer ratio (SJR) in the x-axis of the plot. The rate-reliability with full precoding protects all subcarriers equally and therefore performs better than the rate-reliability beamformer where only a few arbitrary subcarriers are protected. It is evident that both the rate-reliability beamformer designs significantly outperform the benchmark scheme. 

Figure \ref{fig:fp_barr} studies the effect of barrage jamming with varying power levels on the three schemes. Here again, since the barrage jamming affects all the subcarriers, rate-reliability beamformer with full precoding performs better in contrast to rate-reliability with multiprecoding. Both schemes significantly outperform the benchmark scheme by 2 and 3 orders of magnitude respectively.

The effect of multiband disguised jamming is demonstrated in Fig. \ref{fig:fp_mb}. Since only a portion of the spectrum is affected, both rate-reliability schemes perform equally well as all affected subcarriers are protected. Similar to the previous tests, the benchmark scheme fails to perform. 

Finally, to relate the effect of jamming on spectral efficiency, we analyze the spectral efficiency with all three schemes under the most detrimental jamming scheme (disguised all-band). We define spectral efficiency as the number of useful bits transmitted per channel use \cite{anu_mimo} given by
\begin{equation}
    \eta = \mathcal{R}\log_2{\vert \mathcal{Q} \vert}\left( 1-BER\right)
\end{equation}

where $\mathcal{R}$ is the symbol transmission rate and $\vert \mathcal{Q} \vert$ denotes the cardinality of the QAM constellation $\mathcal{Q}$. Both rate-reliability schemes have a symbol transmission rate of $\mathcal{R}=2$ symbols/s whereas the benchmark scheme has a rate of $\mathcal{R}=1$ symbols/s. It is evident that the rate-reliability schemes achieve its full spectral efficiency at SJR = $-10$ dB while the benchmark scheme was still able to achieve only $\sim2.8$ b/s/Hz.





%
%
%

These tests reveal the weaknesses of the conventional Alamouti beamforming scheme and the benefits of adopting the rate-reliability beamforming under active jamming attacks. The reliability of the schemes were demonstrated by the BER performance in Figures \ref{fig:fp_all}, \ref{fig:fp_barr}, and \ref{fig:fp_mb}. The rate performance is depicted by the spectral efficiency gain under the most severe jamming attack in Fig. \ref{fig:mp_mb}.


\section{Conclusion}
\label{sec:con}

In this article, we presented a novel rate-reliability beamformer design and its performance under diverse jamming attacks. We demonstrate how superior rate and reliability performance can be achieved by adopting such jammer-resilient schemes for sustained operation under hostile jamming attacks.  
The most disruptive jamming strategies such as disguised and barrage were considered to validate the jammer resiliency of the proposed schemes in Fig.\ref{fig:fp_all}-Fig.\ref{fig:mp_mb}. The rate-reliability beamformer with multiprecoding demonstrated $2$ orders of magnitude improved BER performance compared to conventional Alamouti beamformer scheme under disguised all-band and barrage jamming. The proposed rate-reliability beamformer with full precoding exhibited $2$ orders of magnitude superior BER performance under all-band and barrage jamming. Under multi-band jamming, the rate-reliability beamformer with full and multiprecoding offered $3$ orders of magnitude higher performance compared to conventional Alamouti beamformer scheme. Finally, the proposed rate-reliability scheme achieved $1.4\times$ higher spectral efficiency in contrast to the conventional scheme under all band jamming. Therefore, the proposed solution offers very high spectral efficiency and reliability under severe jamming attacks while achieving a very low decoding complexity rendering it suitable for practical jammer resilient applications.

\label{sec:refs}

\bibliographystyle{IEEEbib}
\bibliography{strings,refs}
\end{document}